\documentclass[pdflatex,sn-mathphys-num,referee]{sn-jnl}
\usepackage{graphicx}
\usepackage{amsmath,amssymb,mathtools}
\usepackage{booktabs}
\usepackage{float}
\usepackage{microtype}
\usepackage{xcolor}
\usepackage{hyperref}
\hypersetup{hidelinks,hypertexnames=false,pdfauthor={Alejandro Rodriguez Dominguez},pdftitle={Switching Frictions, Heterogeneous Trading Horizons, and Long-Memory Order Flow},pdfsubject={Public preprint, 2 September 2026}}

\theoremstyle{thmstyleone}
\newtheorem{theorem}{Theorem}
\newtheorem{proposition}[theorem]{Proposition}
\newtheorem{corollary}[theorem]{Corollary}
\newtheorem{lemma}[theorem]{Lemma}
\theoremstyle{thmstyletwo}
\newtheorem{assumption}[theorem]{Assumption}

\newcommand{\E}{\mathbb{E}}
\newcommand{\Pp}{\mathbb{P}}
\newcommand{\Cov}{\operatorname{Cov}}
\newcommand{\Var}{\operatorname{Var}}
\newcommand{\dd}{\mathrm{d}}

\begin{document}
\setstretch{1.55}

\title[Switching frictions and long-memory order flow]{Switching Frictions, Heterogeneous Trading Horizons, and Long-Memory Order Flow}

\author*[1,2,3]{\fnm{Alejandro} \sur{Rodr\'iguez Dom\'inguez}}\email{arodriguez@miraltabank.com}
\affil*[1]{\orgdiv{Quantitative Analysis and Artificial Intelligence Department}, \orgname{Miralta Finance Bank S.A.}, \orgaddress{\city{Madrid}, \country{Spain}}}
\affil[2]{\orgdiv{Department of Computer Science}, \orgname{University of Reading}, \orgaddress{\city{Reading}, \country{United Kingdom}}}
\affil[3]{\orgdiv{Department of Data and AI}, \orgname{Albert School}, \orgaddress{\city{Paris}, \country{France}}\\[6pt]\textnormal{2 September 2026}}
\date{2 September 2026}

\abstract{This paper develops a mechanism through which costly changes in the representations used for portfolio choice can contribute to persistent signed order flow. Heterogeneous switching thresholds and opportunity volatility generate heterogeneous residence times, and renewal aggregation maps their execution-weighted tail into the decay of aggregate flow covariance. Under common execution weights, the same tail determines the exponent of representation-spell durations, the order-flow memory exponent, and the horizon at which finite-market scaling must end. First-passage renewal analysis establishes these joint restrictions. Structural simulations recover them from realized paths, quantify the distortion created by mismatched weights, and show how finite cross sections shorten the usable inference horizon. The resulting empirical protocol converts an aggregate persistence fit into cross-dataset restrictions that can determine whether a duration-based kernel is suitable for a separate execution-cost model.}

\maketitle

\section{Introduction}\label{sec:intro}

Signed order flow is one of the most persistent observables in financial markets. Buyer-initiated trades tend to be followed by buyer-initiated trades and seller-initiated trades by seller-initiated trades over horizons far longer than a conventional short-memory benchmark would suggest \cite{LilloFarmer2004,AxioglouSkouras2011}. The leading structural explanation is order splitting: institutions execute large parent orders gradually, and heterogeneous metaorder sizes or durations generate persistent trade signs \cite{LilloMikeFarmer2005,VaglicaLilloMoroMantegna2008,MoroEtAl2009,TothPalitLilloFarmer2015,SatoKanazawa2023}. This paper asks whether a complementary source of persistence can arise upstream, when institutions decide whether to retain or replace the representation that organizes portfolio choice.

The paper forms the temporal and microstructure component of a representation-based research programme. Rodríguez Domínguez \cite{RodriguezDominguez2026b} studies portfolio policy when the common-driver geometry changes; Rodríguez Domínguez \cite{RodriguezDominguez2026Representation} studies equilibrium selection, crowding, cross-impact, and capacity under shared representations; and Rodríguez Domínguez \cite{RodriguezDominguez2026Stability} develops inference and capacity control when the resulting feedback boundary must be estimated. The present paper supplies the link from representation choice to market observables by deriving how replacement costs, residence heterogeneity, and execution timing jointly restrict representation-spell durations, order-flow persistence, and the finite-market cutoff.

Portfolio decisions depend on a predictive representation: a factor set, common-driver subspace, model family, or information state that organizes signals, forecasts, constraints, and hedges. Replacing that representation is costly. Models must be re-estimated and validated, exposures and hedges must change, and a new target portfolio must be implemented. These costs create an inaction region. A portfolio retains its incumbent representation until the advantage of an alternative is large enough to justify switching. The maintained representation need not fix the sign of every trade. Instead, it may induce a persistent conditional component of desired demand, around which news, rebalancing, and execution noise can produce sign changes within the same spell.

The analysis links three layers of the mechanism. A local reflected-diffusion approximation turns switching frictions into a residence-time scale. Renewal aggregation then maps the cross-sectional tail of that scale into the decay rate of aggregate signed demand, and asynchronous execution determines the weights with which portfolios enter observed market flow. A finite market cannot possess a literal power law at every horizon: in an exact truncated-Pareto benchmark, the observable scaling window ends near the square of the upper residence scale.

This structure produces observable restrictions rather than only an aggregate fit. A sufficient product-tail condition links heterogeneity in switching costs and opportunity volatility to the residence-scale tail. The same tail determines both spell-duration tails and the decay of execution-weighted order-flow covariance. The upper residence scale determines where the intermediate scaling range must end. Dependence, a heavier inverse-volatility tail, or a different execution clock can change these mappings, so the maintained conditions remain visible throughout the analysis.

The paper contributes a switching-friction foundation for a duration-driven component of order flow and derives the joint restrictions needed to test it. Heterogeneous aggregation and duration-driven dependence are established ideas \cite{Granger1980,TaqquLevy1986,Parke1999,HsiehHurvichSoulier2007}; the new step is to tie their mixing distribution to portfolio inaction, determine the observational weights from execution, and require the same tail to predict spell durations, flow persistence, and the end of the scaling range. This converts a flexible aggregate fit into a set of cross-dataset restrictions. The contribution is therefore not another explanation of how duration mixtures can generate slow decay. It is the economic origin, execution-weighted observation rule, and joint duration--flow--cutoff restriction that can reject the proposed channel. The duration restriction permits a direct comparison between observed representation spells and held-out flow decay without recovering every latent residence scale.

The present mechanism is self-contained: it does not require the equilibrium results of the companion papers, and it does not treat every order-flow episode as a representation switch.

Aggregate autocorrelation alone cannot identify the channel. Heavy-tailed metaorders, slowly changing regimes, and self-exciting flow can generate similar persistence \cite{DieboldInoue2001,BacryMuzy2014,JaissonRosenbaum2016}. A decisive test therefore requires portfolio or account residence spells jointly observed with signed executions and must retain explanatory power after reconstructed-metaorder and common-shock controls. The duration tail and flow covariance must also be estimated under the same execution weights. These requirements distinguish a test of representation residence from a comparison between unrelated persistent series.

For practical use, the paper separates validation from optimization. The empirical protocol begins with a defensible representation measure and a dated record of representation changes. It then applies the same execution weights to residence spells and signed flow, estimates duration tails on a training sample, and evaluates unconditional aggregate-flow covariance and its cutoff on held-out observations. Within-spell comparisons are used only to test alignment, not to estimate the memory exponent. A kernel satisfying these restrictions can enter a separate impact-and-cost model; it does not itself determine an order, schedule, or capacity limit.

Section 2 relates the mechanism to aggregation, market microstructure, and inaction. Section 3 develops the model and its observable restrictions. Section 4 reports computational evidence and the empirical implementation. Section 5 concludes.

\section{Related literature and positioning}\label{sec:literature}

Cross-sectional aggregation can turn heterogeneous short-memory components into long memory \cite{Granger1980}. Renewal and stochastic-duration constructions generate related effects when sufficiently long episodes receive non-negligible probability \cite{TaqquLevy1986,Parke1999,HsiehHurvichSoulier2007}. Regular variation supplies the transfer tools used below \cite{BinghamGoldieTeugels1987}, while classical renewal identities provide the stationary residual-life calculation \cite{Cox1962,Feller1971}. The distinction here is economic and observational: duration heterogeneity comes from portfolio switching scales, and the relevant cross section is weighted by executed contribution rather than by a simple count of agents.

Long-memory trade signs are extensively documented \cite{LilloFarmer2004}. The order-splitting literature derives persistence from heavy-tailed hidden-order sizes and durations, reconstructs institutional metaorders, and tests the implied exponent relation \cite{LilloMikeFarmer2005,VaglicaLilloMoroMantegna2008,MoroEtAl2009,TothPalitLilloFarmer2015,SatoKanazawa2023}. Heterogeneous participation changes amplitudes and the effective population weights without necessarily changing the leading exponent \cite{SatoKanazawa2024}. Optimal execution provides the economic reason why desired positions are implemented gradually \cite{Kyle1985,BertsimasLo1998,AlmgrenChriss2001,ObizhaevaWang2013}. The present mechanism can coexist with splitting because representation residence is upstream of, and typically longer than, any one metaorder.

Metaorder reconstruction also clarifies what aggregate decay cannot identify. Parent-order persistence is supported when reconstructed order size or duration predicts subsequent trade signs \cite{VaglicaLilloMoroMantegna2008,MoroEtAl2009,TothPalitLilloFarmer2015}. A representation-residence mechanism instead requires the maintained portfolio state to predict signed flow beyond the life of individual parent orders. The two channels therefore call for different observational units and different held-out controls, even when they produce similar unconditional autocorrelations.

Self-exciting models provide a different route to persistent market activity \cite{BacryMuzy2014,JaissonRosenbaum2016}. Recent unified models jointly study order flow, volatility, and impact \cite{MuhleKarbeEtAl2026,MaitrierBouchaud2025}. No impact equation is imported here: doing so would add a model-dependent restriction without strengthening identification of the switching channel.

Non-convex adjustment costs produce inaction and heterogeneous adjustment hazards \cite{BertolaCaballero1990,CaballeroEngel1999}. Fixed transaction costs similarly lead to optimal stopping in portfolio problems \cite{MortonPliska1995}, while slow-moving institutional capital affects prices and flows \cite{Duffie2010,VayanosWoolley2013}. Representation-based portfolio choice supplies the economic object that is maintained between switches \cite{RodriguezDominguez2026b,RodriguezDominguez2026Representation}. Relative to that work, this paper isolates the timing of representation replacement and its microstructure implications. Relative to the order-flow literature, it contributes execution-aligned restrictions on durations, covariance decay, and the finite-market cutoff. It does not solve a full representation equilibrium or claim that the proposed channel dominates order splitting.

Gebbie \cite{Gebbie2026Hierarchical} distinguishes event dynamics, actor-conditioned admissibility, and the clock that embeds events in calendar time. The model below selects a single first-passage renewal law and an independent execution clock, then derives quantitative duration--flow restrictions under that selection. A state-dependent or non-unique clock would change the calendar-time mapping and lies outside the execution-clock invariance result; the paper therefore does not claim to identify the wider hierarchical causal structure.

Table \ref{tab:channels} makes the identifying distinction explicit. Order splitting and representation residence can coexist, but they attach persistence to different units and therefore imply different held-out comparisons. A fit to aggregate flow alone cannot distinguish them.

\begin{table}[t]
\centering
\caption{Observable distinctions among persistence channels.}
\label{tab:channels}
\small
\begin{tabular}{p{.16\textwidth}p{.20\textwidth}p{.27\textwidth}p{.22\textwidth}}
\toprule
Channel & Persistent unit & Restriction beyond aggregate decay & Binding control \\
\midrule
Order splitting & Parent order or metaorder & Metaorder size or duration predicts trade-sign persistence & Reconstructed parent orders \\
Representation residence & Portfolio representation spell & Execution-weighted spell tail predicts flow decay and its cutoff & Same-account spells, signed executions, and aligned weights \\
Common regime & Market-wide state & Persistence survives without account-specific spell alignment & Common-shock and break controls \\
Self-excitation & Event history & Conditional intensity explains dependence after observed state controls & Residual event-time diagnostics \\
\bottomrule
\end{tabular}
\end{table}

\section{Switching, aggregation, and executed order flow}\label{sec:model}

\subsection{Switching times and residence scales}

The timing argument starts from standard first-passage and stationary-renewal identities for reflected Brownian motion \cite{Cox1962,Feller1971}. These identities fix the residence scale that later enters the observable duration and flow restrictions.

Consider portfolios indexed by $a$. Portfolio $a$ uses an incumbent predictive representation and faces an effective switching threshold $\kappa_a>0$. Let $G_{a,t}\geq0$ denote a one-dimensional local opportunity coordinate measuring accumulated evidence in favor of replacing the incumbent representation. Between switches,
\begin{equation}
 \dd G_{a,t}=\sigma_a\dd W_{a,t}+\dd L^0_{a,t},\qquad G_{a,t}\geq0,
 \label{eq:reflected}
\end{equation}
where $\sigma_a>0$, $W_a$ is standard Brownian motion, and $L^0_a$ is local time enforcing reflection at zero. A switch occurs at
\begin{equation}
 \tau_a=\inf\{t>0:G_{a,t}=\kappa_a\}.
\end{equation}
After switching, the coordinate is reset and a new spell begins.

Equation \eqref{eq:reflected} is a declared local approximation. It is not the maximum of many unrestricted continuation-value processes and does not claim that all representation choice is one-dimensional. Its role is to isolate a falsifiable scale restriction. Reflected Brownian motion started at zero has the law of absolute Brownian motion, hence
\begin{equation}
 \tau_a\overset{d}{=}R_a^2\tau_0,
 \qquad R_a:=\frac{\kappa_a}{\sigma_a},
 \label{eq:scale}
\end{equation}
where $\tau_0$ is the first exit time of standard Brownian motion from $(-1,1)$ and $\E\tau_0=1$. Thus $R_a$ has units of square-root time and $\E\tau_a=R_a^2$.

The threshold $\kappa_a$ is reduced form but economically interpretable. It is the local advantage required to offset implementation, validation, governance, funding, and turnover costs. The paper does not infer these components separately. Its testable object is their ratio to the rate at which replacement opportunities arrive, $R_a=\kappa_a/\sigma_a$. This distinction prevents the first-passage benchmark from being read as a complete optimal-stopping solution. A fully structural stopping model would be needed to decompose $\kappa_a$ or evaluate counterfactual policy changes; it is not needed for the downstream duration--flow restriction once the scale law in \eqref{eq:scale} is maintained.

The Laplace transform and survival function are
\begin{align}
 \E[e^{-s\tau_a}]&=\operatorname{sech}(R_a\sqrt{2s}), \label{eq:laplace}\\
 \Pp(\tau_0>t)&=\frac4\pi\sum_{n=0}^{\infty}\frac{(-1)^n}{2n+1}
 \exp\!\left[-\frac{(2n+1)^2\pi^2}{8}t\right]. \label{eq:survival}
\end{align}
These formulas make the renewal clock non-exponential at finite lags, while preserving exact scale homogeneity.

\subsection{Aggregation and observable restrictions}\label{sec:aggregation}

The effective threshold may aggregate implementation, turnover, model replacement, validation, governance, and funding frictions. Linking its cross-sectional distribution to opportunity volatility provides a transparent route from those primitives to a heavy residence-scale tail and, in turn, to two separately observable objects.

\begin{proposition}[A sufficient primitive tail condition]\label{prop:breiman}
Let $\kappa>0$ and $\sigma>0$ denote a random switching threshold and opportunity volatility, and set $R=\kappa/\sigma$. Suppose $\kappa$ is regularly varying with index $\theta>0$,
\begin{equation}
 \Pp(\kappa>x)=x^{-\theta}L_\kappa(x),
\end{equation}
$\kappa$ and $\sigma$ are independent, and $\E[\sigma^{-(\theta+\varepsilon)}]<\infty$ for some $\varepsilon>0$. Then
\begin{equation}
 \Pp(R>r)\sim \E[\sigma^{-\theta}]\Pp(\kappa>r),
 \qquad r\to\infty.
 \label{eq:primitive}
\end{equation}
Consequently, $R$ has tail index $\theta$.
\end{proposition}

This is an application of Breiman's product lemma to $\kappa\sigma^{-1}$ \cite{Breiman1965}. It is sufficient, not necessary. If inverse volatility is heavier-tailed than the switching threshold, or if the two variables are strongly dependent, the tail of $R$ can instead be governed by those features. The proposition therefore converts an otherwise reduced-form tail assumption into a concrete empirical diagnostic without pretending to derive a universal equilibrium distribution.

Conditional on $R_a$, let representation spells have iid durations $R_a^2\tau_0$. Spell $n$ has representation state $S_{a,n}$ and a centered conditional-demand component
\begin{equation}
 B_{a,n}:=\E[D_a(t)\mid S_{a,n}],\qquad
 \E B_{a,n}=0,\qquad \Var(B_{a,n})=v_a\in(0,\infty).
 \label{eq:emission}
\end{equation}
The $B_{a,n}$ are iid across spells and independent of durations. Within a spell,
\begin{equation}
 D_a(t)=B_{a,n(t)}+\varepsilon_a(t),
 \label{eq:demand}
\end{equation}
where $\varepsilon_a$ is centered, stationary, independent of the renewal process and spell states, and has absolutely integrable (or, in discrete time, summable) covariance $\gamma_{\varepsilon,a}(t)$. Thus the representation affects the conditional mean of desired signed demand; it does not impose a constant trade sign. The process is initialized in its stationary renewal law.

\begin{lemma}[Stationary renewal kernel \cite{Cox1962,Feller1971}]\label{lem:kernel}
For two times separated by $t\geq0$, the covariance of the persistent component is
\begin{equation}
 \Cov(B_{a,n(0)},B_{a,n(t)}\mid R_a)=v_a h_0(t/R_a^2),
 \label{eq:kernel}
\end{equation}
where
\begin{equation}
 h_0(x)=\int_x^\infty\Pp(\tau_0>u)\dd u
 =\frac{32}{\pi^3}\sum_{n=0}^{\infty}\frac{(-1)^n}{(2n+1)^3}
 \exp\!\left[-\frac{(2n+1)^2\pi^2}{8}x\right].
 \label{eq:hseries}
\end{equation}
In particular, $h_0(0)=1$, $h_0$ is positive, and it decays exponentially.
\end{lemma}

A renewal between the observation times refreshes the centered conditional-demand component and kills its covariance. The remaining probability is the equilibrium residual-life survival function, which yields \eqref{eq:kernel}. Consequently,
\begin{equation}
 \Cov(D_a(0),D_a(t)\mid R_a)=v_a h_0(t/R_a^2)+\gamma_{\varepsilon,a}(t).
 \label{eq:demandcov}
\end{equation}
No forced sign reversal, constant-sign spell, or exponential-sojourn approximation is used.

Let $b_a$ be the contribution scale of portfolio $a$ and define
\begin{equation}
 X_N(t)=\sum_{a=1}^N b_aD_a(t),\qquad
 \mu_N(\dd r)=\frac{\sum_{a=1}^N b_a^2v_a\delta_{R_a}(\dd r)}{\sum_{a=1}^N b_a^2v_a}.
 \label{eq:latentmeasure}
\end{equation}
Assume independence across portfolios. After normalization by $\sum_a b_a^2v_a$, aggregate covariance is a mixture of $h_0(t/R_a^2)$ with respect to $\mu_N$, plus the normalized transitory-demand covariance
\begin{equation}
 s_N(t):=\frac{\sum_a b_a^2\gamma_{\varepsilon,a}(t)}{\sum_a b_a^2v_a}.
 \label{eq:shortremainder}
\end{equation}

\begin{assumption}[Weighted residence tail]\label{ass:rv}
The measures $\mu_N$ converge weakly to a probability law $\mu$ satisfying
\begin{equation}
 \overline\mu(r):=\mu((r,\infty))\sim L(r)r^{-\theta},
 \qquad \theta>0,
 \label{eq:rv}
\end{equation}
where $L$ is slowly varying. Moreover, $s_N(t)\to s(t)$ pointwise and
\begin{equation}
 s(t)=o\!\left(L(\sqrt t)t^{-\theta/2}\right).
 \label{eq:shortmemory}
\end{equation}
\end{assumption}

The limiting normalized covariance is
\begin{equation}
 C_\infty(t)=\int_0^\infty h_0(t/r^2)\mu(\dd r)+s(t).
 \label{eq:mixture}
\end{equation}

\begin{theorem}[Residence-tail mapping]\label{thm:aggregation}
Under Assumption \ref{ass:rv},
\begin{equation}
 C_\infty(t)\sim K_{\mathrm{FP}}(\theta)L(\sqrt t)t^{-\theta/2},
 \qquad
 K_{\mathrm{FP}}(\theta)=\frac{\theta}{2}\int_0^\infty h_0(u)u^{\theta/2-1}\dd u.
 \label{eq:main}
\end{equation}
The constant is finite and strictly positive. Hence the aggregate memory exponent is $\alpha=\theta/2$.
\end{theorem}

\begin{corollary}[Memory boundary]\label{cor:boundary}
For integer sampling, covariance is non-summable when $0<\theta<2$ and summable when $\theta>2$. In the former case,
\begin{equation}
 \Var\!\left(\sum_{t=1}^{T}X(t)\right)\asymp T^{2-\theta/2}L(\sqrt T).
 \label{eq:variance}
\end{equation}
At $\theta=2$, non-summability holds if and only if
\begin{equation}
 \sum_{k\geq1}\frac{L(\sqrt k)}{k}=\infty.
 \label{eq:boundary}
\end{equation}
\end{corollary}

The boundary statement matters: a tail index alone does not determine the critical case. Equation \eqref{eq:variance} is a variance-growth result, not by itself a functional convergence theorem.

\begin{corollary}[Observable spell-duration restriction]\label{cor:duration}
Suppose the execution-weighted residence scale has tail $L_F(r)r^{-\theta}$ and the first-passage duration satisfies $\tau=R^2\tau_0$, with $\tau_0$ independent of $R$. Then
\begin{equation}
 \Pp(\tau>t)\sim \E[\tau_0^{\theta/2}]L_F(\sqrt t)t^{-\theta/2}.
 \label{eq:durationtail}
\end{equation}
Thus the execution-weighted spell-duration tail exponent $\beta$ and the order-flow memory exponent obey $\alpha_F=\beta=\theta/2$.
\end{corollary}

This corollary replaces a latent-scale comparison with an observable duration restriction. It is valid only when spell durations and order flow use the same execution weights; an unweighted duration sample generally targets a different cross section.

\subsection{Finite markets and executed flow}\label{sec:execution}

Every fixed finite mixture is ultimately short-memory because its residence scales are bounded. To characterize the transition, consider the normalized Pareto law truncated at $x>1$,
\begin{equation}
 \mu_x(\dd r)=\frac{\theta r^{-\theta-1}\mathbf 1_{[1,x]}(r)}{1-x^{-\theta}}\dd r,
 \qquad
 C_x(t)=\int h_0(t/r^2)\mu_x(\dd r).
 \label{eq:truncated}
\end{equation}

\begin{theorem}[Truncated-Pareto window and cutoff]\label{thm:finite}
Let $C_\infty$ be the same mixture under the untruncated Pareto law. For sequences $t\to\infty$ and $x\to\infty$:
\begin{enumerate}
 \item if $t/x^2\to0$, then $C_x(t)/C_\infty(t)\to1$;
 \item if $t/x^2\to\infty$, then for finite constants $A,\kappa>0$ independent of $x$ and $t$,
 \begin{equation}
  C_x(t)\leq \frac{A}{1-x^{-\theta}}\exp(-\kappa t/x^2).
  \label{eq:postcutoff}
 \end{equation}
\end{enumerate}
Moreover, in the first regime the relative truncation error is
\begin{equation}
 \left|\frac{C_x(t)}{C_\infty(t)}-1\right|
 =O\!\left(x^{-\theta}+\left(\frac{\sqrt t}{x}\right)^\theta\right).
 \label{eq:truncerror}
\end{equation}
\end{theorem}

Thus, in this benchmark, the square of the truncation scale separates a regime asymptotically equivalent to the infinite-mixture power law from a regime with an exponential upper bound. If $R_1,\ldots,R_N$ are iid Pareto, extreme-value theory separately gives
\begin{equation}
 N^{-1/\theta}R_{\max,N}\Rightarrow Z_\theta,
 \qquad R_{\max,N}^2=O_p(N^{2/\theta}),
 \label{eq:max}
\end{equation}
where $Z_\theta$ is Fr\'echet. The resulting stochastic cutoff proxy grows as $N^{2/\theta}$. This order statement does not turn a random finite sample into the deterministic truncated-Pareto law. Genuine long memory belongs to the cross-sectional limit; a finite market can display only an intermediate scaling window.

Replacing the first-passage durations by exponential clocks with the same mean would preserve the leading exponent but change the kernel and prefactor. The exact clock is therefore economically relevant at finite horizons even when a Markov approximation is exponent-correct.

Observed order flow weights residence spells by their executed contribution. Let $N_a(t)$ be an independent Poisson process of execution opportunities with intensity $\lambda_a>0$, and let $q_a>0$ be quantity per execution. Cumulative signed flow is
\begin{equation}
 F_N(t)=\sum_{a=1}^N q_a\int_0^tD_a(s-)\dd N_a(s).
 \label{eq:flow}
\end{equation}
For bins of width $\Delta$ and nonzero lag $\ell$,
\begin{align}
 \Cov(F_{N,0}^{\Delta},F_{N,\ell}^{\Delta})
 =\sum_{a=1}^Nq_a^2\lambda_a^2\int_0^\Delta\!\int_0^\Delta
 \left\{v_a h_0\!\left(\frac{\ell\Delta+v-u}{R_a^2}\right)
 +\gamma_{\varepsilon,a}(\ell\Delta+v-u)\right\}\dd u\dd v.
 \label{eq:flowcov}
\end{align}
The Poisson martingale affects contemporaneous variance but not covariance between disjoint bins. Define the execution-weighted measure
\begin{equation}
 \mu_N^F(\dd r)=
 \frac{\sum_aq_a^2\lambda_a^2v_a\delta_{R_a}(\dd r)}{\sum_aq_a^2\lambda_a^2v_a}.
 \label{eq:qmeasure}
\end{equation}

\begin{theorem}[Execution-clock invariance]\label{thm:execution}
Suppose $\mu_N^F$ converges to a law with tail $L_F(r)r^{-\theta}$. Assume the normalized execution-weighted contribution of $\gamma_{\varepsilon,a}$ converges to a remainder that is $o(L_F(\sqrt t)t^{-\theta/2})$. Take the cross-sectional limit first. For fixed $\Delta>0$,
\begin{equation}
 \Gamma_\Delta^F(\ell)\sim
 \Delta^2K_{\mathrm{FP}}(\theta)L_F(\sqrt{\ell\Delta})(\ell\Delta)^{-\theta/2},
 \qquad \ell\to\infty,
 \label{eq:execution}
\end{equation}
where $\Gamma_\Delta^F$ is covariance normalized by $\sum_aq_a^2\lambda_a^2v_a$.
\end{theorem}

Participation heterogeneity and the variance of representation-conditioned demand jointly determine which portfolios dominate the measured tail. An unweighted estimate of residence scales generally tests the wrong restriction. Poisson execution is a transparent benchmark; arbitrary dependent clocks are not covered. A short-memory arrival process should preserve the result only under explicit mixing, moment, and independence conditions. If arrivals themselves have long memory, their contribution must be identified jointly.

\section{Computational evidence and empirical implementation}\label{sec:numerics}

\subsection{Structural validation and operational variants}

The experiments simulate realized first-passage renewals and asynchronous executions rather than fitting curves generated directly from the limiting formulas. They remain structural validation, not market evidence. Each renewal process is initialized from its equilibrium residual distribution; portfolio-level spell states are then observed only at independent Poisson execution times. Thirty replications are used for each of three residence-tail indices, with 500 portfolios and 4,096 observation periods per replication; an independent sample of 30,000 durations is used for each spell-tail estimate. The exit-clock draws have mean 1.001 and variance 0.668, while the equilibrium residual draws have mean 0.833. The complete sampling and estimation sequence is specified in the appendix.

Figure \ref{fig:structural-recovery} compares the target half-index restriction with estimates from realized spell durations and executed-flow covariance. The duration estimates recover the targets closely. Flow estimates are centered near the restriction for target exponents 0.400 and 0.600, with means 0.409 and 0.603 and Monte Carlo standard errors 0.009 and 0.014. At the 0.800 target, the mean rises to 0.898 with a Monte Carlo standard error of 0.041; its 10th--90th percentile range is [0.670, 1.217]. The corresponding duration estimate is 0.801 with a Monte Carlo standard error of 0.003. The deterioration is informative for practice: near the short-memory boundary, this flow estimator needs a longer horizon or an explicitly bias-corrected design, and the spell-tail estimate is materially more stable.

\begin{figure}[t]
\centering
\includegraphics[width=.94\textwidth]{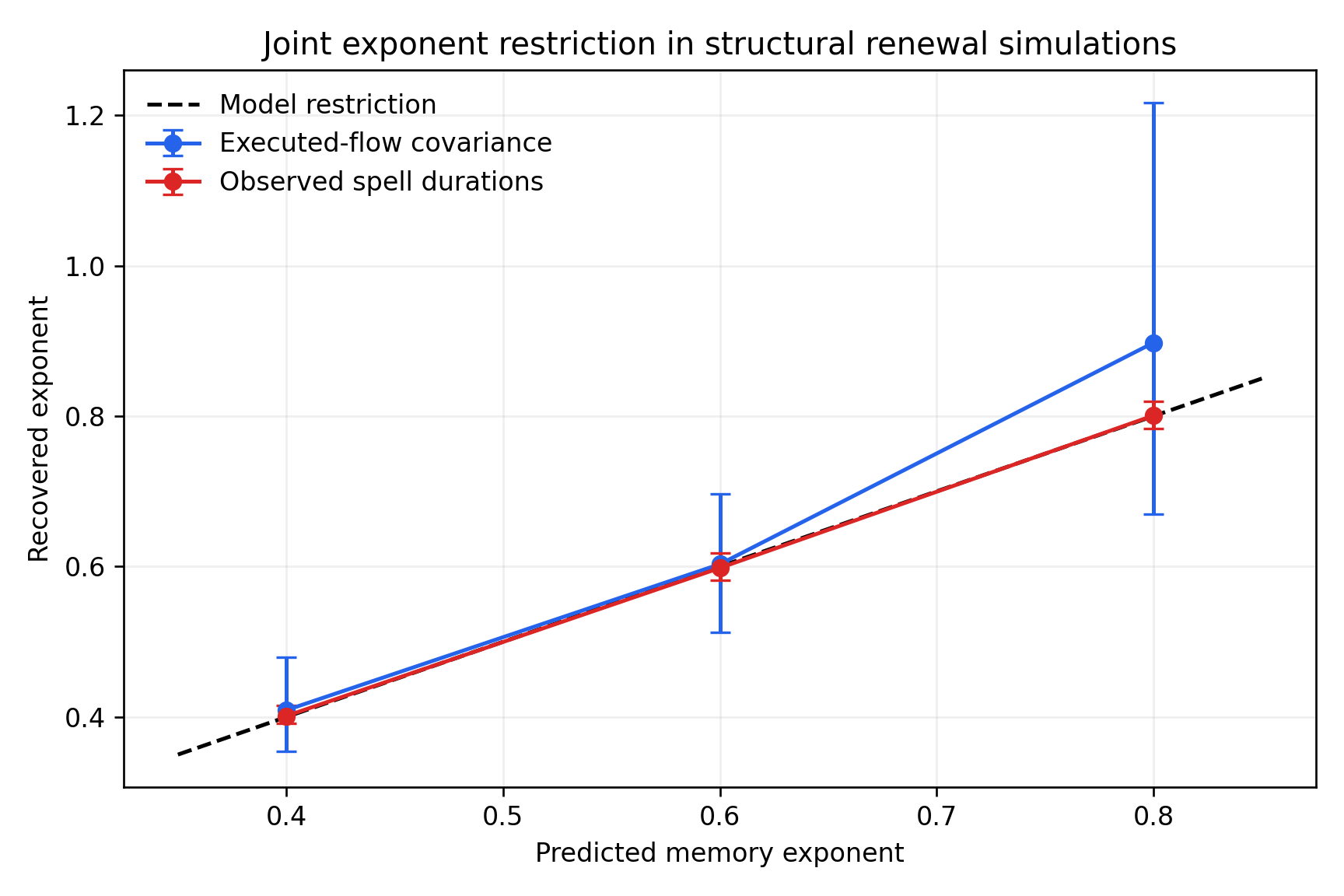}
\caption{Joint exponent restriction in structural renewal simulations. Points summarize estimates from realized spell paths and independently sampled executions; error bars show the 10th--90th percentile range across replications. The diagonal is the analytical restriction, not a fitted regression.}
\label{fig:structural-recovery}
\end{figure}

\begin{figure}[t]
\centering
\includegraphics[width=.94\textwidth]{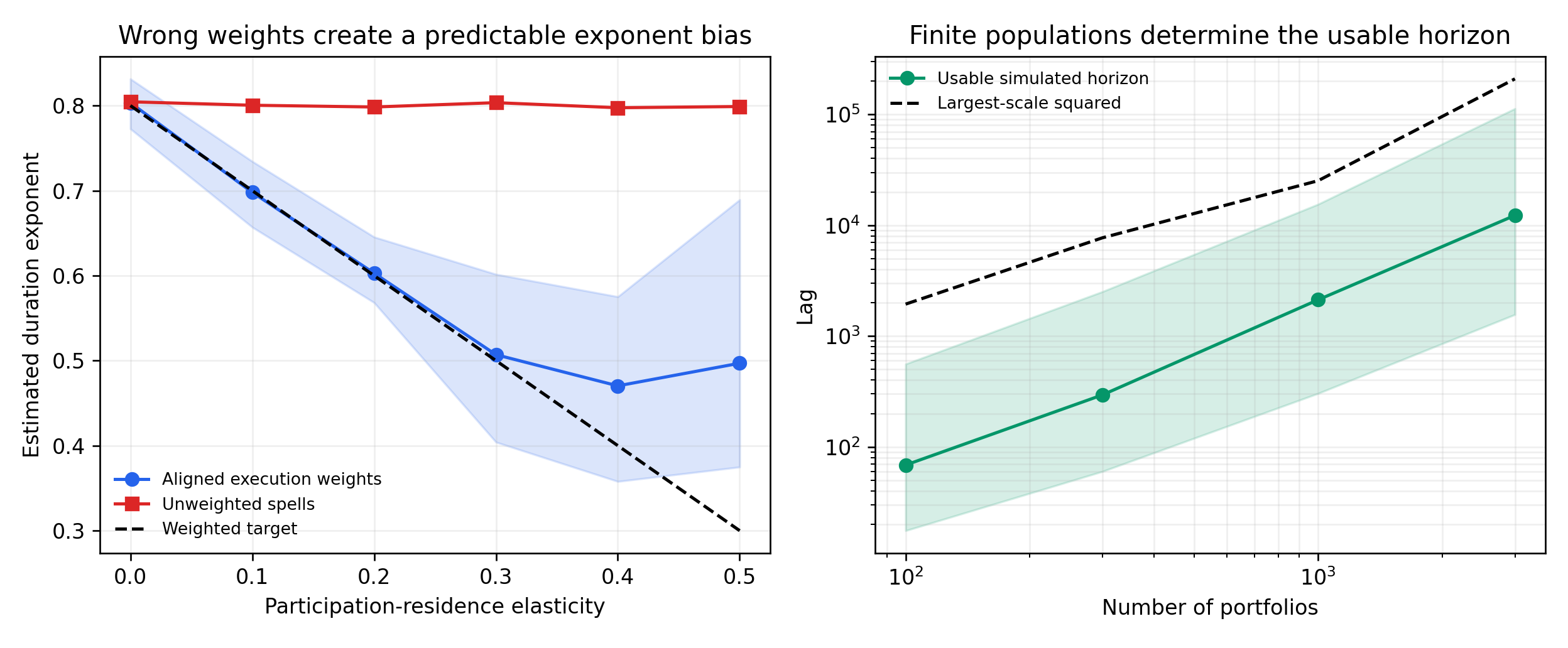}
\caption{Operational variants. Left: execution-aligned and unweighted spell-tail estimates as participation becomes more concentrated in long-residence portfolios. Right: empirical usable scaling horizon across finite cross sections; bands show the 10th--90th percentile range.}
\label{fig:operational-variants}
\end{figure}

The left panel of Figure \ref{fig:operational-variants} changes the observational problem rather than the theorem. Execution contribution is allowed to rise with residence scale, so an unweighted spell sample and the execution-weighted flow target different cross sections. Aligned weighting tracks the changed target until extreme concentration leaves too little effective sample, whereas the unweighted estimate stays near 0.800 even when the relevant target falls. At participation elasticities from 0 to 0.300, aligned mean estimates move from 0.803 to 0.507 as the target moves from 0.800 to 0.500; unweighted means remain between 0.797 and 0.804. At larger elasticities the aligned estimator becomes unstable, which should trigger an effective-sample-size warning rather than a forced exponent claim.

The right panel asks how far a finite cross section supports a stable scaling interpretation. Across 60 draws per population size, the median usable horizon rises from 68.539 periods at 100 portfolios to 12,270.751 at 3,000 portfolios, but dispersion is wide. The realized stable range is consistently shorter than the crude largest-residence-scale proxy. Practitioners should therefore estimate the cutoff from stability across lag windows and cross sections, using the largest observed residence only as an upper-scale diagnostic.

Exact-series, exponential-clock, quadrature, and deterministic truncated-Pareto comparisons provide lower-level numerical checks. They support the sampling construction but are not used as the paper's main experimental evidence.

\subsection{Empirical design and interpretation}\label{sec:identification}

A direct test needs account or portfolio identifiers, holdings or target positions, signed executions, and a defensible reconstruction of the representation used by each account. Versioned production models provide the cleanest representation record. When the state must be estimated, a switch should be recorded only when the inferred representation changes beyond its sampling uncertainty, and uncertain change dates should be treated as interval-censored. Residence spells then measure switching horizons, while realized squared quantity, participation intensity, and the variance of representation-conditioned demand construct the weights in \eqref{eq:qmeasure}.

The empirical design must separate estimation and evaluation samples. The training sample estimates the tail index of the execution-weighted residence distribution or spell durations. The evaluation sample estimates signed-flow decay without reselecting the lag window to match the prediction. The primary restrictions are
\begin{equation}
 H_0^{(R)}:\quad 2\alpha_F-\theta=0,
 \qquad
 H_0^{(\tau)}:\quad \alpha_F-\beta=0.
 \label{eq:test}
\end{equation}
The duration restriction applies only when the duration and flow samples use the same execution weights. The flow exponent must be estimated from unconditional aggregate covariance: conditioning both observations on membership in the same spell removes the survival probability that generates the decay in Theorem \ref{thm:aggregation}. Same-spell and different-spell dependence remain useful as an auxiliary alignment comparison, but they do not estimate $\alpha_F$. A further restriction compares the end of the observed scaling range with the estimated upper residence scale across institutions, markets, or subsamples. Joint agreement is more informative than fitting aggregate autocorrelation alone.

Metaorder splitting is the principal confound. The test should reconstruct parent orders, estimate persistence within them, and repeat the analysis on residual signed flow. Common information shocks and mechanically synchronized portfolio rules require time and asset controls. Structural breaks can mimic long memory in finite samples \cite{DieboldInoue2001,AxioglouSkouras2011}; tail estimates, lag-window choices, and break placebos must therefore be reported together.

The proposed channel is empirically admissible only if the held-out exponent relation is supported, residence scales predict the finite-population cutoff, and the relation remains after metaorder and common-shock controls. A lack of portfolio-level representation data cannot be converted into evidence for or against the mechanism. Aggregate fit is at most compatibility evidence.

The local coordinate in \eqref{eq:reflected} may be inadequate when switching opportunities are multidimensional, have material drift, jump, or exhibit state-dependent volatility. Dependence between switching costs, opportunity volatility, representation-conditioned demand, and execution can also alter the tail or covariance mapping. These are substantive failures, not technical details. Proposition \ref{prop:breiman} and Theorems \ref{thm:finite}--\ref{thm:execution} make the needed conditions visible so that each can be tested or relaxed separately.

The theory concerns one source of persistence and does not claim exclusivity. Order splitting and self-excitation may remain quantitatively dominant. The useful question is whether representation residence contributes incremental, out-of-sample explanatory power at horizons longer than individual metaorders.

Together, these restrictions define the empirical conditions in Table
\ref{tab:operational}. They prevent a persistence estimate from entering an
execution-cost model when the duration and flow objects are not observationally
aligned. The conditions govern the admissibility of the estimated kernel; they
do not generate trades or an execution schedule.

\begin{table}[t]
\centering
\caption{Conditions for an empirically admissible persistence kernel.}
\label{tab:operational}
\renewcommand{\arraystretch}{1.15}
\begin{tabular}{p{.20\textwidth}p{.34\textwidth}p{.34\textwidth}}
\toprule
Condition & Required implementation & Interpretation \\
\midrule
Measurement & Dated representation states, account identifiers, signed executions, parent-order links, and pre-specified controls & Spells and flow refer to the same decision unit \\[4pt]
Weighting & Apply the same estimates of $q_a^2\lambda_a^2v_a$ to durations and covariance & The two exponents are comparable under Corollary \ref{cor:duration} \\[4pt]
Held-out tests & Use unconditional flow covariance to test the duration exponent and cutoff on an evaluation sample & The joint restrictions, rather than an aggregate fit, determine consistency \\[4pt]
Alignment & Compare within-spell dependence with cross-spell and time-shifted benchmarks after metaorder and common-shock controls & Tests incremental representation content without redefining the memory exponent \\
\bottomrule
\end{tabular}
\end{table}

\section{Conclusion}\label{sec:conclusion}

Switching frictions can transform heterogeneous portfolio inaction into persistent aggregate order flow when representation spells carry a nonzero persistent component of desired demand. The exact first-passage renewal clock yields a half-index mapping from the execution-weighted residence tail to both the spell-duration tail and the order-flow memory exponent. A primitive product-tail condition identifies one economic route to that tail. Independent asynchronous execution preserves the exponent, and the truncated-Pareto benchmark locates the end of the intermediate scaling window.

The empirical contribution is the joint test implied by the model. Portfolio residence spells and held-out signed flow must use the same execution weights, the flow exponent must be estimated from unconditional covariance, and the upper residence scale must explain the end of the scaling range as well as its slope. This design separates the proposed channel from an unrestricted long-memory fit and from persistence created by order splitting or common shocks. When the restrictions are supported, the resulting kernel can inform a subsequent execution-cost model while remaining distinct from the trading decision itself.

\section*{Declarations}
\textbf{Funding.} No external funding was received. 
\textbf{Competing interests.} The author is employed by a financial institution that trades asset classes related to the subject of this study. The views expressed are the author's own. 
\textbf{Data availability.} No proprietary or client data are used. The numerical evidence is generated from the structural designs and sampling rules described in the appendix.

\appendix
\section{Proofs for the switching and renewal results}\label{app:renewal}

\subsection{First-passage formulas}

Because reflected Brownian motion started at zero has the law of absolute Brownian motion, its hitting time of $R$ is the first exit time from $(-R,R)$. For $u(x)=\E_x[e^{-s\tau_R}]$, the generator equation is
\begin{equation}
 \tfrac12u''(x)=su(x),\qquad u(-R)=u(R)=1.
\end{equation}
Symmetry gives $u(x)=\cosh(\sqrt{2s}x)/\cosh(\sqrt{2s}R)$ and hence \eqref{eq:laplace}. Brownian scaling yields \eqref{eq:scale}. The eigenfunction expansion gives \eqref{eq:survival}. Integrating it term by term and using $\E\tau_0=1$ gives \eqref{eq:hseries}.

For a waiting time $\tau$ with finite mean, the equilibrium residual-life survival is
\begin{equation}
 \Pp(R_e>t)=\frac1{\E\tau}\int_t^\infty\Pp(\tau>u)\dd u.
\end{equation}
With iid centered conditional-demand components, a renewal between the observation times makes their product have zero expectation. Their conditional covariance is therefore $v_a$ times the residual survival. Independence of $\varepsilon_a$ adds $\gamma_{\varepsilon,a}(t)$, proving Lemma \ref{lem:kernel} and \eqref{eq:demandcov}.

\subsection{Proof of Proposition \ref{prop:breiman}}

Write $Y=\sigma^{-1}$. Then $R=\kappa Y$, $\kappa$ is regularly varying with index $\theta$, and $\E[Y^{\theta+\varepsilon}]<\infty$. Breiman's lemma implies
\begin{equation}
 \Pp(\kappa Y>r)\sim\E[Y^\theta]\Pp(\kappa>r),
\end{equation}
which is \eqref{eq:primitive}.

\subsection{Proof of Theorem \ref{thm:aggregation}}

Set $T=R^2$ and $\alpha=\theta/2$. Assumption \ref{ass:rv} implies
\begin{equation}
 \Pp(T>x)\sim L(\sqrt x)x^{-\alpha}.
\end{equation}
Since $h_0(y)=\int_y^\infty\Pp(\tau_0>u)\dd u$, Tonelli's theorem yields
\begin{equation}
 \E[h_0(t/T)]=\int_0^\infty\Pp(\tau_0>u)\Pp(T\geq t/u)\dd u.
 \label{eq:tonelli}
\end{equation}
For fixed $u>0$, the integrand after normalization by $L(\sqrt t)t^{-\alpha}$ converges to $\Pp(\tau_0>u)u^\alpha$. Potter bounds dominate the normalized tail by a constant times $u^{\alpha-\varepsilon}+u^{\alpha+\varepsilon}$. Brownian exit survival is bounded at zero and exponentially decreasing at infinity, so this envelope is integrable. The domain where $t/u$ remains bounded begins at a multiple of $t$ and contributes an exponentially small remainder. Dominated convergence in \eqref{eq:tonelli} gives
\begin{equation}
 \frac{t^\alpha}{L(\sqrt t)}\E[h_0(t/T)]
 \longrightarrow\int_0^\infty\Pp(\tau_0>u)u^\alpha\dd u.
\end{equation}
Integration by parts, with $h_0'(u)=-\Pp(\tau_0>u)$, gives the constant in \eqref{eq:main}. Positivity and finiteness follow from positivity and exponential decay of $h_0$. The remainder $s(t)$ is lower order by \eqref{eq:shortmemory}, so it does not alter the equivalent.

For Corollary \ref{cor:boundary}, Karamata's theorem for sums gives the variance order in \eqref{eq:variance} when $0<\alpha<1$. When $\alpha>1$, covariance is summable. At $\alpha=1$, comparison with $L(\sqrt k)/k$ gives \eqref{eq:boundary}.

For Corollary \ref{cor:duration}, the execution-weighted variable $R^2$ is regularly varying with index $\theta/2$. The exit time $\tau_0$ has moments of every positive order because its survival function decays exponentially. Breiman's product lemma applied to $R^2\tau_0$ gives \eqref{eq:durationtail}. Theorem \ref{thm:execution} then gives the same exponent for execution-weighted order-flow covariance.

\section{Proof for the finite-population benchmark}\label{app:finite}

For the Pareto density, direct substitution $u=t/r^2$ gives
\begin{equation}
 C_\infty(t)=\frac{\theta}{2}t^{-\theta/2}
 \int_0^t h_0(u)u^{\theta/2-1}\dd u.
 \label{eq:paretoexact}
\end{equation}
The integral converges to a positive finite constant. Let
\begin{equation}
 I_x(t)=\int_1^x h_0(t/r^2)\theta r^{-\theta-1}\dd r.
\end{equation}
Then $C_x(t)=I_x(t)/(1-x^{-\theta})$ and
\begin{equation}
 0\leq C_\infty(t)-I_x(t)
 =\int_x^\infty h_0(t/r^2)\theta r^{-\theta-1}\dd r
 \leq x^{-\theta}.
 \label{eq:missingtail}
\end{equation}
By \eqref{eq:paretoexact}, $C_\infty(t)$ is asymptotic to a positive constant times $t^{-\theta/2}$. Dividing \eqref{eq:missingtail} by that quantity gives a relative missing-tail error of order $(\sqrt t/x)^\theta$. Normalization contributes
\begin{equation}
 \left|(1-x^{-\theta})^{-1}-1\right|=O(x^{-\theta}).
\end{equation}
This proves \eqref{eq:truncerror} and the first claim.

From \eqref{eq:hseries}, there are finite $A,\kappa>0$ such that $h_0(y)\leq A e^{-\kappa y}$ for all $y\geq0$. For $r\leq x$, $h_0(t/r^2)\leq A e^{-\kappa t/x^2}$. Integration with respect to the unnormalized truncated Pareto density and division by $1-x^{-\theta}$ gives \eqref{eq:postcutoff}. Finally,
\begin{equation}
 \Pp(N^{-1/\theta}R_{\max,N}\leq z)
 =\left(1-\frac{z^{-\theta}}N\right)^N\longrightarrow e^{-z^{-\theta}},
\end{equation}
which proves \eqref{eq:max}.

\section{Execution-clock covariance}\label{app:execution}

Write $N_a(t)=\lambda_at+\widetilde N_a(t)$. Compensated Poisson increments over disjoint bins are orthogonal. Conditional on the desired-demand path,
\begin{equation}
 \E[F_{a,k}^{\Delta}\mid D_a]
 =q_a\lambda_a\int_{k\Delta}^{(k+1)\Delta}D_a(t)\dd t.
\end{equation}
Taking covariance across disjoint bins and using stationarity gives \eqref{eq:flowcov}. Dividing by the total persistent execution contribution converts the cross-sectional sum into integration with respect to \eqref{eq:qmeasure}; the transitory term is lower order by assumption.

For each fixed lag, boundedness and continuity of the bin-integrated kernel permit the cross-sectional limit. Theorem \ref{thm:aggregation} then supplies a regularly varying limiting covariance. Because $v-u$ ranges over a compact interval, the uniform convergence theorem for regularly varying functions gives
\begin{equation}
 \frac{C_F(\ell\Delta+v-u)}{C_F(\ell\Delta)}\longrightarrow1
\end{equation}
uniformly over the bin. Integration contributes $\Delta^2$ and proves \eqref{eq:execution}. The theorem takes the cross-sectional limit before the long-lag limit; the reverse order is not asserted.

\section{Simulation and replication protocol}\label{app:simulation}

The structural experiments use random seed 260902. First-exit times are drawn
from the exponential-product representation of
\(\operatorname{sech}(\sqrt{2s})\), using 48 explicit components and a
moment-matched gamma remainder. Equilibrium residual times are obtained from
the corresponding length-biased distribution. Pools of 200,000 first-exit and
200,000 residual draws are generated before the experiments begin.

For the exponent-recovery experiment, repeat the following procedure 30 times
for each \(\theta\in\{0.8,1.2,1.6\}\):

\begin{enumerate}
\item Draw 500 residence scales from the Pareto law
\(\Pr(R>r)=r^{-\theta}\), \(r\ge1\).
\item Initialize every portfolio with an equilibrium residual spell, assign an
independent centered binary spell state, and simulate renewals for 4,096
periods using durations \(R^2\tau_0\).
\item Draw independent unit-rate Poisson execution counts and aggregate the
executed signed states. Estimate the covariance-decay exponent by log--log
regression over positive covariance estimates at lags 24--256.
\item Independently draw 30,000 residence scales and first-exit times, form the
spell durations, and estimate their tail exponent with a Hill estimator using
the largest 8\% of observations.
\end{enumerate}

The execution-weighting experiment fixes \(\theta=1.6\), draws 15,000
residence scales per replication, and repeats each participation elasticity
60 times. Execution intensity is \(R^\eta\exp(0.2Z)\), where \(Z\) is standard
normal, so the aligned estimator uses squared intensity as its weight. The
finite-cross-section experiment fixes \(\theta=1.4\), considers
\(N\in\{100,300,1000,3000\}\), and uses 60 independent cross sections per
population size. Its usable horizon is the last point in the first run of
local log-slope estimates that remains within 0.20 of \(\theta/2\), allowing
five consecutive observations before entry or exit. These definitions produce
the summaries in Figures \ref{fig:structural-recovery} and
\ref{fig:operational-variants}.

\section{Empirical implementation protocol}\label{app:empirical-protocol}

An empirical application should first define the observational unit and the
representation maintained by that unit. Production-model identifiers,
approved factor sets, or dated investment-process states are preferable to
portfolio-composition proxies. When the representation is inferred rather
than recorded, its uncertainty and the interval containing each change date
must be retained when spells are constructed.

The duration and flow samples must then be placed on a common observational
measure. For each account, estimate executed quantity, participation
intensity, and the variance of representation-conditioned demand, and use the
resulting estimate of $q_a^2\lambda_a^2v_a$ in both samples. Split accounts or
non-overlapping time blocks into training and evaluation sets before selecting
tail estimators, lag ranges, or cutoff rules. The training sample estimates
the weighted residence-scale or spell-duration tail and the upper residence
scale. The evaluation sample aggregates signed executions across accounts and
estimates unconditional covariance at the pre-specified lags. Pairs must not
be restricted to observations within the same spell when estimating the
memory exponent.

Inference should report the held-out differences $2\alpha_F-\theta$ and
$\alpha_F-\beta$, together with a comparison between the observed end of the
scaling range and the predicted residence-scale cutoff. Account-clustered
resampling preserves within-account spells; calendar blocks preserve common
market dependence. Both are needed when the application contains many
accounts and a shorter common time series.

The final stage evaluates identification rather than re-estimating the
exponent. Reconstructed parent orders remove the component attributable to
order splitting, time and asset effects absorb common shocks, and structural
break diagnostics distinguish regime mixtures from long memory. Within-spell
dependence can be compared with cross-spell and time-shifted benchmarks, but
this is an alignment test only. A persistence kernel is carried into a
separate impact-and-cost model only when the weighted duration, unconditional
flow, cutoff, and alignment results support the same mechanism.

\bibliography{references}

\end{document}